\documentclass[aps,prb,twocolumn,superscriptaddress]{revtex4-2}

\usepackage[colorlinks=true,citecolor=blue]{hyperref}
\usepackage{graphicx}
\usepackage{amsmath}
\usepackage{amssymb}
\usepackage{bbold}
\usepackage{braket}

\DeclareGraphicsExtensions{.png,.jpg,.eps}
\usepackage{xcolor}

\begin{document}

\author{Tania Paul}
\affiliation{International Research Centre MagTop, Institute of Physics, Polish Academy of Sciences, Aleja Lotnik\'ow 32/46, PL-02668 Warsaw, Poland}

\author{V. Fern\'andez Becerra}
\affiliation{International Research Centre MagTop, Institute of Physics, Polish Academy of Sciences, Aleja Lotnik\'ow 32/46, PL-02668 Warsaw, Poland}
\author{Timo Hyart}
\affiliation{International Research Centre MagTop, Institute of Physics, Polish Academy of Sciences, Aleja Lotnik\'ow 32/46, PL-02668 Warsaw, Poland}
\affiliation{Department of Applied Physics, Aalto University, 00076 Aalto, Espoo, Finland}

\title{Interplay of quantum spin Hall effect and spontaneous time-reversal symmetry breaking in electron-hole bilayers I: Transport properties}

\date{\today}
   
\begin{abstract}
The band-inverted electron-hole bilayers, such as InAs/GaSb, are an interesting playground for the interplay of quantum spin Hall  effect and correlation effects because of the small density of electrons and holes and the relatively small hybridization between the electron and hole bands. It has been proposed that Coulomb interactions lead to a time-reversal symmetry  broken phase when the electron and hole densities are tuned from the trivial to the quantum spin Hall insulator regime. We show that the transport properties of the system in the time-reversal symmetry broken phase are consistent with the recent experimental observations in InAs/GaSb. Moreover, we carry out a quantum transport study on a Corbino disc where the bulk and edge contributions to the conductance can be separated.  We show that the edge becomes smoothly conducting and the bulk is always  insulating when one tunes the system from the trivial to the quantum spin Hall insulator phase, providing unambiguous transport signatures of the time-reversal symmetry broken phase.
\end{abstract}

\maketitle

\section{Introduction}

The advent of topological materials \cite{RevKane,RevZhang} has brought band-inverted semiconductors, with small electron and hole densities, to the focus of the attention in the search of quantum spin Hall (QSH) insulators \cite{BHZ,Liu08,QSH_HgTe, Knez11, Du15, Wu18}. However, the electron-electron interactions are  important in these materials if the hybridization of the electron and hole bands is small compared to the exciton binding energy, as can be appreciated by noticing that the bilayer system of spatially separated electrons and holes is the well-known paradigm system for the realization of an exciton condensate state  \cite{lozovik1976new, Naveh96}. Indeed, it is now theoretically understood that interactions can lead to a plethora of correlated phases in band-inverted semiconductors \cite{Pikulin14, Budich14, Hu17, Xue18, Zhu19, Varsano20, zeng2021inplane} and the recent experiments have shown evidence of excitonic phenomenology in InAs/GaSb quantum wells \cite{Du-exciton, Wu19, Du19PRB, Xiao19, Irie-gap20} as well as in WTe$_2$ \cite{WTe2-exciton, WTe2-exciton2}.

\begin{figure}
    \centering
    \includegraphics[width=0.83\columnwidth]{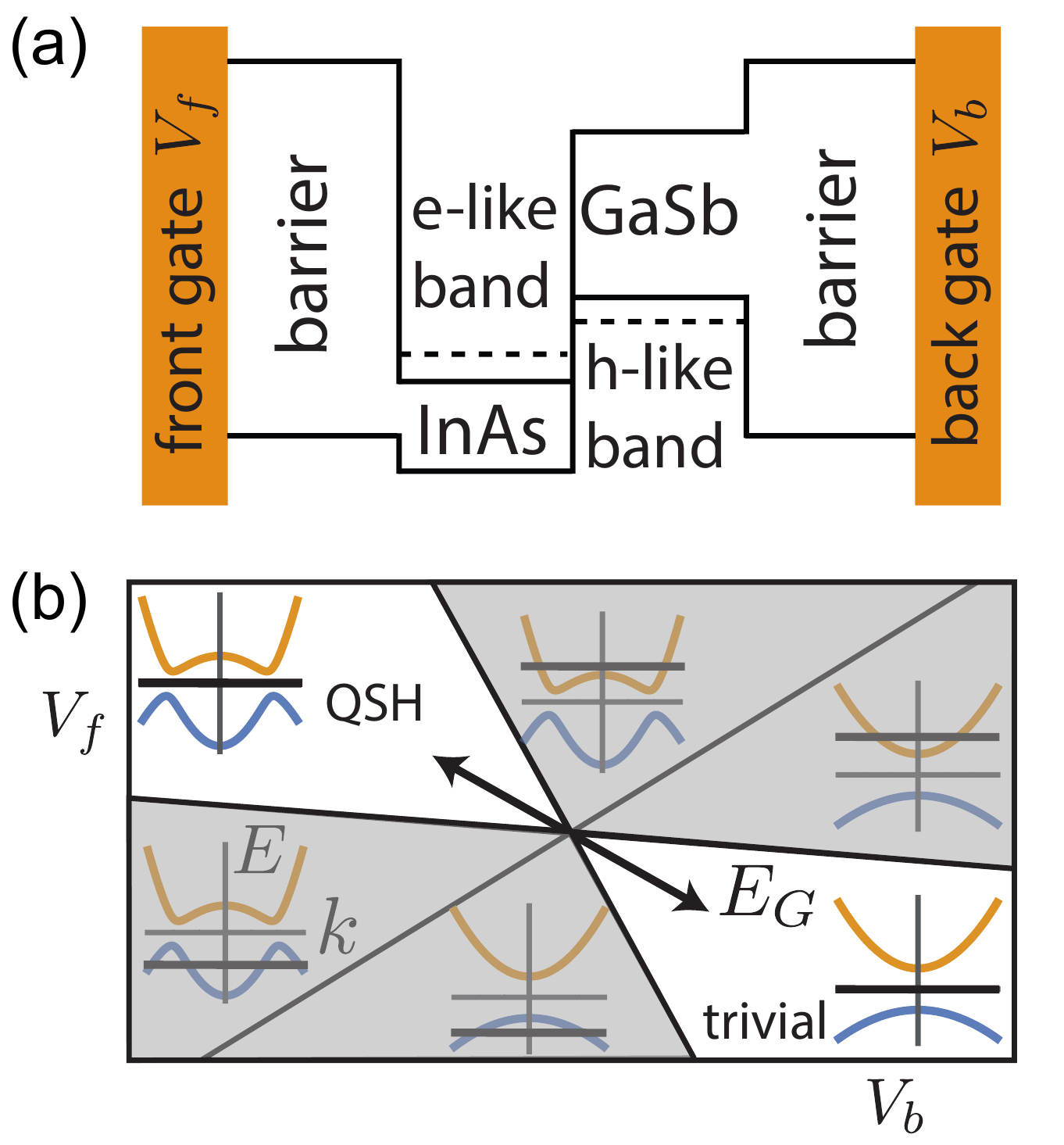}
	\caption{Schematic illustration of the setup. (a) The densities of the  electrons and holes can be controlled with gate voltages $V_f$ and $V_b$ in a heterostructure supporting spatially separated electron and hole bands.  (b) This way the gate voltages determine whether the electron and hole bands are inverted at the $\Gamma$ point ($E_G>0$) or not ($E_G<0$), as well as whether the Fermi level (thick black line) is in the conduction band, band gap or valence band.  The insulating phase with $E_G>0$ ($E_G<0$) is the QSH (trivial) insulator phase.}
    \label{fig:schematic}
\end{figure}
We concentrate on the correlated phases appearing in the band-inverted electron-hole bilayers shown in Fig.~\ref{fig:schematic}(a) \cite{Liu08}. In these systems, the electron and hole bands are spatially separated and therefore only weakly hybridized. Moreover, the electron and hole densities (and hence also the band-inversion parameter $E_G$) can be controlled {\it in situ} with front and back  gate voltages, $V_f$ and $V_b$, allowing the possibility to study the phase transition between  trivial and QSH insulator phases \cite{Liu08, PhysRevLett.115.036803, Irie-gap20}, as schematically illustrated in Fig.~\ref{fig:schematic}(b). It has been theoretically predicted that, due to the excitonic correlations caused by the Coulomb interactions, a third phase with spontaneously broken time-reversal symmetry (TRS) will appear in the transition regime between the two topologically distinct phases \cite{Pikulin14}. Within this phase the helical edge states, originating from the QSH insulator phase, can exist but they are not protected against backscattering, and it was theoretically demonstrated \cite{Pikulin14} that these unprotected edge states can explain the  temperature-independent mean free path observed in InAs/GaSb bilayers in the presence of reasonably large applied currents  \cite{Knez14, Spanton14, Du15}. However, an unambiguous experimental demonstration of the existence of the exotic insulating phase with spontaneously broken TRS symmetry is still lacking in these systems. 

Here, we demonstrate that the transport properties of the system in the TRS broken phase are also consistent with the more recent transport experiments in InAs/GaSb bilayers with small applied currents \cite{Du-transport}, so that the spontaneous TRS symmetry breaking provides a comprehensive explanation of the temperature, voltage and length dependencies of the observed conductance \cite{Knez14, Spanton14, Du15, Du-transport}. Finally, we propose an experiment which can be used to unambiguously demonstrate the existence of the spontaneous TRS breaking in this system. Namely, we show that the edge becomes smoothly conducting and the bulk remains insulating when one tunes across the TRS broken phase appearing between the trivial and QSH insulator phases in the Corbino geometry, where the bulk and edge contributions to the conductance can be separated 
 \cite{Corbino-decoupling}. In the presence of TRS symmetry the bulk transport gap must close when the system is tuned between topologically distinct phases, and hence the experimental demonstration of a transition without a bulk transport gap closing constitutes a proof of an existence of TRS broken insulating phase.

\section{Spontaneous TRS breaking in electron-hole bilayers}

In Ref.~\onlinecite{Pikulin14} it was shown using a full Hartree-Fock calculation that the Coulomb interactions in the Bernevig-Hughes-Zhang (BHZ) model \cite{BHZ} developed for InAs/GaSb bilayers \cite{Liu08, liu2013models} lead to three different phases as a function of the hybridization of the electron and hole bands $A$ and the band-inversion parameter $E_G$, which is defined here so that for $E_G>0$ ($E_G<0$) the electron and hole bands are (not) inverted at the $\Gamma$ point, see Fig.~\ref{fig:schematic}(b). As intuitively expected, for small (large) $A$ and $E_G$ one realizes a trivial (QSH) insulator phase. However, interestingly it was found that at intermediate values of $A$ and $E_G$ there exists an insulating phase with spontaneously broken TRS symmetry separating the topologically distinct phases. In this Section we describe a simplified {\it minimal model} that fully captures all the essential results obtained using the full Hartree-Fock calculations in Ref.~\onlinecite{Pikulin14}.

The single particle BHZ Hamiltonian is
\begin{equation}
H_0 =  \big(E_G - \frac{\hbar^2 k^2}{2m}\big) \tau_z \sigma_0    + A k_x \tau_x \sigma_z   - A k_y \tau_y \sigma_0 + \Delta_{z} \tau_y  \sigma_y,  \label{eq:BHZ}
\end{equation}
where $\tau$'s and $\sigma$'s denote the Pauli matrices in the electron-hole and spin  basis, respectively. 
The electron band is made out of $s$-orbitals and the hole band is made out of only two $p$-orbitals, because the electric confining potential and the atomic spin-orbit coupling remove the degeneracies of the $p$-orbitals.  The tunneling between the layers is dominantly odd in momentum and opens up a hybridization gap $\propto A$. Here, we have assumed the same effective mass $m$ for electrons and holes, and included only the momentum-independent spin-orbit coupling term $\Delta_z$ arising due to bulk inversion asymmetry. 
We have ignored the asymmetry of the masses and the momentum-dependent spin-orbit coupling terms, because they are not essential for understanding the phase diagram of the InAs/GaSb bilayers \cite{Pikulin14}.

The main effect of Coulomb interactions 
is the binding of the electrons and holes into excitons with the characteristic size $d_0$ and binding energy $E_0$ determined by the relation 
$
E_0=\hbar^2/ (m d_0^2)=e^2/(4\pi \epsilon \epsilon_0d_0).
$
This leads to an excitonic mean field \cite{Pikulin14}
\begin{eqnarray}
 H_{EC}&=&  \Re[\Delta_1] \tau_y \sigma_y +\Re[\Delta_2] \big[k_x \tau_x \sigma_z  -  k_y \tau_y \sigma_0  \big] \nonumber\\ && + \Im[\Delta_1] \tau_x  \sigma_y -\Im[\Delta_2] \big[k_x \tau_y \sigma_z  + k_y \tau_x \sigma_0   \big],  \label{ec-mean}
\end{eqnarray}
where  $\Delta_1$ and $\Delta_2$ are complex bosonic fields describing $s$-wave and $p$-wave excitonic correlations,   respectively. For simplicity we have expanded the fields $\Delta_1$ and $\Delta_2$ only to the lowest order in momentum and neglected the full  $|\mathbf{k}|$ dependence, which is present in the numerical solution  of the Hartree-Fock equations \cite{Pikulin14}. It is easy to see by straightforward calculation that the terms in the first line of Eq.~(\ref{ec-mean}) obey the TRS $T=i\tau_0\sigma_y K$  ($K$ is the complex conjugation operator) and  the terms in the second line break it. Therefore, the imaginary parts of the fields $\Im[\Delta_1], \Im[\Delta_2] \ne 0$ result in spontaneous TRS breaking.

\begin{figure}
    \centering
    \vspace{0.08cm}
    \includegraphics[width=0.782\linewidth]{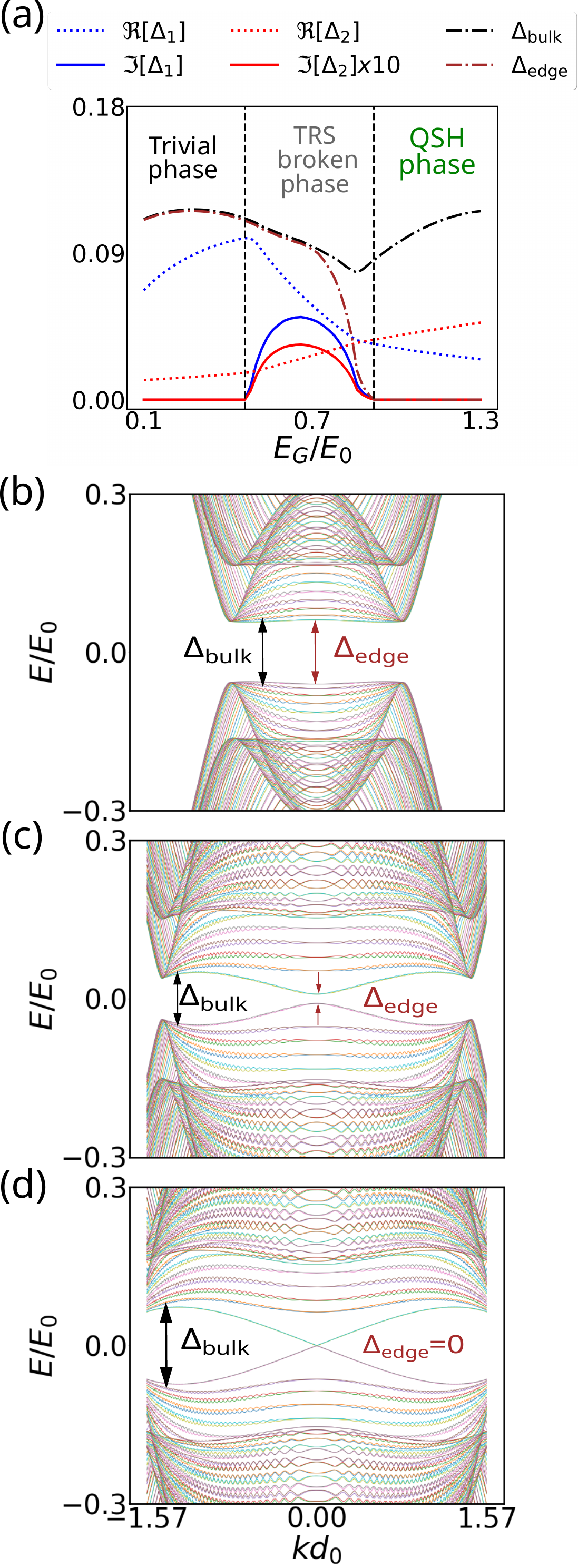}
	\caption{(a) Phase diagram as a function of $E_G$. The trivial and QSH phases obey the TRS. In the TRS broken phase  the $s$- and $p$-wave excitonic mean fields obey $\Im[\Delta_1], \Im[\Delta_2] \ne 0$. The bulk gap $\Delta_{\rm bulk}$ remains open for all values of $E_G$ and the edge gap $\Delta_{\rm edge}$ decreases monotonously from the bulk gap value to zero, when one tunes $E_G$ across the TRS broken phase towards the QSH phase. The model parameters are described in the text. Energy bands in (b) trivial phase with $E_G=0.3 E_0$, (c) TRS broken phase with $E_G=0.86E_0$ and (d) QSH phase with $E_G=1.12 E_0$. The eigenenergies are obtained by diagonalizing the tight-binding Hamiltonian which is generated from the continuum Hamiltonian,  defined by Eqs.~(\ref{eq:BHZ})-(\ref{p-wave-mean}), using the Kwant software package \cite{Groth_2014}.
}
    \label{fig:delta_bulk_gap}
\end{figure}

We can solve the complex bosonic mean fields $\Delta_1$ and $\Delta_2$ by substituting  the ansatz (\ref{ec-mean}) to the Hartree-Fock mean field equations. This way, we arrive to the following mean field equations (see Appendix \ref{APP:A} for more details)
\begin{equation}
\Delta_1=\frac{g_s d_0^2}{(2\pi)^2} \int d^2 k \ \big[\langle c_{\mathbf{k}\downarrow2}^\dag c_{\mathbf{k}\uparrow 1} \rangle-\langle c_{\mathbf{k}\uparrow2}^\dag c_{\mathbf{k}\downarrow 1} \rangle \big] \label{s-wave-mean}
\end{equation}
and
\begin{eqnarray}
\Delta_2&=&\frac{g_p d_0^4}{(2\pi)^2} \int d^2 k \ \big[ -\langle c_{\mathbf{k}\uparrow2}^\dag c_{\mathbf{k}\uparrow 1} \rangle(k_x-ik_y) \nonumber \\ && \hspace{2.2cm}+\langle c_{\mathbf{k}\downarrow2}^\dag c_{\mathbf{k}\downarrow 1} \rangle (k_x+ik_y) \big], \label{p-wave-mean}
 \end{eqnarray}
 where $g_s$ ($g_p$) is the effective interaction strength for $s$-wave ($p$-wave) pairing and $c_{1 \sigma k}$ ($c_{2 \sigma k}$) is the electron annihilation operator  with spin $\sigma$ and momentum $k$ in electron (hole) layer. In our numerical calculations the integration is performed over the range $|\mathbf{k}| \leq 2.26/d_0$, but the exact values of the integration limits are not important. The effective interaction strengths $g_s$ and $g_p$ can be considered as fitting parameters, whose values should be fixed so that one approximately reproduces the results obtained from Hartree-Fock calculations \cite{Pikulin14}.

The values of the model parameters for InAs/GaSb can be estimated by combining theoretical calculations \cite{Liu08, liu2013models, Naveh96, Pikulin14} and the experimentally observed energy gaps \cite{Du15, Du-exciton}. This way, we arrive to parameter values that are used in our  calculations: $E_0/k_B = 200$ K, $d_0=10$ nm,  $A/(E_0d_0)=0.06$,  $\Delta_z/E_0=0.02$, $g_s/E_0=1.0$ and $g_p/E_0=0.2$. The band-inversion parameter $E_G$ is a gate-tunable parameter (see Fig.~\ref{fig:schematic}), which is varied in our calculations to tune the system from trivial insulator to QSH insulator phase. As shown in Fig.~\ref{fig:delta_bulk_gap}, our simplified mean field approach, defined by Eqs.~(\ref{eq:BHZ})-(\ref{p-wave-mean}), reproduces the results obtained from full Hartree-Fock calculations \cite{Pikulin14}. For small (large) values of $E_G$ the system is in a trivial (QSH) insulator phase, and importantly these two phases are separated from each other by an insulating phase with spontaneously broken TRS, where $\Im[\Delta_1], \Im[\Delta_2] \ne 0$. The bulk gap $\Delta_{\rm bulk}$ remains open for all values of $E_G$, because the intermediate TRS broken phase  enables the connection of the topologically distinct phases without bulk gap closing. The edge gap $\Delta_{\rm edge}$ decreases monotonously when one starts from the trivial phase and tunes the system across the TRS broken phase to the QSH phase, where the gapless edge excitations are protected by the topology.

The appearance of spontaneous TRS breaking can be understood with the help of topological considerations. The topological invariant distinguishing the QSH phase from the trivial insulator can change only if
(i) {\it the bulk energy gap closes} or (ii) {\it TRS is broken in a regime between the topologically distinct phases}. The case (i) would be the only possibility if the local order were fixed. However, in an interacting system the order parameter corresponds to a minimum of the free energy, and it is energetically favourable to keep the system gapped. Due to this reason there is a general tendency for the appearance of a TRS broken phase in the transition regime between QSH and trivial insulator phases.

\section{Length, temperature and voltage dependence of the conductance}

The identification of the edge states in InAs/GaSb bilayers was initially problematic due to finite bulk density of states in the minigap \cite{Knez11}.  The main breakthrough in eliminating the bulk conduction came from insertion of Si to the interface between the InAs and GaSb layers during the growth process   \cite{Du15}. 
After achieving a truly insulating bulk this way, L.~Du {\it et al.} \cite{Du15} managed to demonstrate in mesoscopic samples wide conductance plateaus  quantized to the values expected for nonlocal helical edge transport (variations less than $1 \%$).
 The accurate conductance quantization was reported for several devices of various lengths and three different geometries in Ref.~\onlinecite{Du15}.  Moreover, by imaging the distribution of the current flow inside the sample it has been confirmed that the current flows along the edge in agreement with helical edge conduction \cite{Spanton14}. More careful measurements of temperature and voltage dependencies are also consistent with single-mode edge conduction \cite{Du-transport}.  In a different type of samples, where Si was not inserted and the observed thermal activation gap for the bulk transport is an order of magnitude smaller, multi-mode edge conduction has been reported by another group \cite{Nichele_2016}. The explanation of the remarkably different transport properties observed in the presence and in the absence of Si doping remains an open theoretical problem. Because these observations are mutually inconsistent, it is clear that they cannot be explained with the same model Hamiltonian.  Here, we concentrate on the transport experiments in Si doped samples with large activation gap \cite{Du15, Du-transport}. We show that these experiments are consistent with the transport properties theoretically obtained in the TRS broken phase.

In long samples the conductance is not observed to be quantized  \cite{Du15} indicating that backscattering processes occur between the counterpropagating edge channels. It was found that in the limit $eV \gg k_BT$ the resistance is independent on temperature between $20$ mK - $4.2$ K and it increases linearly with the edge length $L$. These observations are not surprising once the elastic backscattering processes are allowed and large voltage is applied, because under these conditions the inelastic scattering rate is expected to be approximately equal to the elastic one  \cite{Bagrets09}, and therefore the localization effects can be neglected and the resistance is expected to be temperature independent.  In the QSH phase the elastic  backscattering is forbidden in the presence of time-reversal symmetry due to the topological protection, so these observations are not consistent with the system being in the QSH phase without additional assumption about the existence of charge puddles that may lead to enhanced backscattering rate \cite{Vayrynen14}. On the other hand, TRS broken phase supports edge states but the elastic backscattering is now allowed, so the experimental observations are fully consistent with the system being in the TRS broken phase. Thus, TRS broken phase provides an intrinsic explanation of these experiments, remaining applicable even if we assume that the samples are of high quality  so that no charge puddles are present in the system.

In short mesoscopic samples with small applied voltage and temperature, the voltage and temperature dependencies of the conductance are more complicated and we need to use a quantum transport approach to describe them. The disorder-averaged differential conductance $G_d=dI/dV$ is obtained from
\begin{equation}
G_d(E_F+eV, T) =  \int_{-\infty}^{+\infty} dE \frac{2 G_0 \ {\rm exp}[-L/\ell(E)]}{4 k_B T \cosh^2 \frac{E - E_F - eV}{2 k_B T}}, \label{Gd_eq}
\end{equation}
where $G_0=e^2/h$, $E_F$ is the Fermi energy, $V$ is voltage, $T$ is temperature of the reservoirs, $L$ is the length of the sample and $\ell(E)$ is the energy-dependent elastic mean free path, which for $E \gg  \Delta_{\rm edge}$ is given by \cite{Pikulin14}
\begin{equation}
\ell(E) = \frac{4 a \hbar^2 v^2 E^2}{\xi V_{\rm dis}^2 \Delta_{\rm edge}^2}. \label{ell_eq}
\end{equation}
Here, $E$ is the energy relative to the energy of the crossing of the edge states, $v$ is the edge velocity, $V_{\rm dis}$ is the strength of the disorder potential, $\xi$ is the disorder correlation length and $a\sim 1$ is a numerical factor. Although the exact expression for $\ell(E)$ is model dependent, it must always satisfy $\ell(E) \to \infty$ for $E \gg \Delta_{\rm edge}$, so that $G_d \approx 2G_0$ for $k_BT \gg \Delta_{\rm edge}$. Therefore, there exists robust asymptotic limits 
\begin{equation}
G_d \approx \begin{cases}
2 G_0 \big[1-L/\ell(E_F+eV)\big], &  k_B T \ll E_F+eV, \\ 
2 G_0, & k_BT \gg \Delta_{\rm edge},
\end{cases}
\end{equation}
which guarantee that $G_d$ undergoes a crossover from non-quantized value to the quantized value $G_d=2G_0$ both with increasing temperature and voltage.

\begin{figure}
\hspace{-0.2cm}\includegraphics[width=0.8 \linewidth]{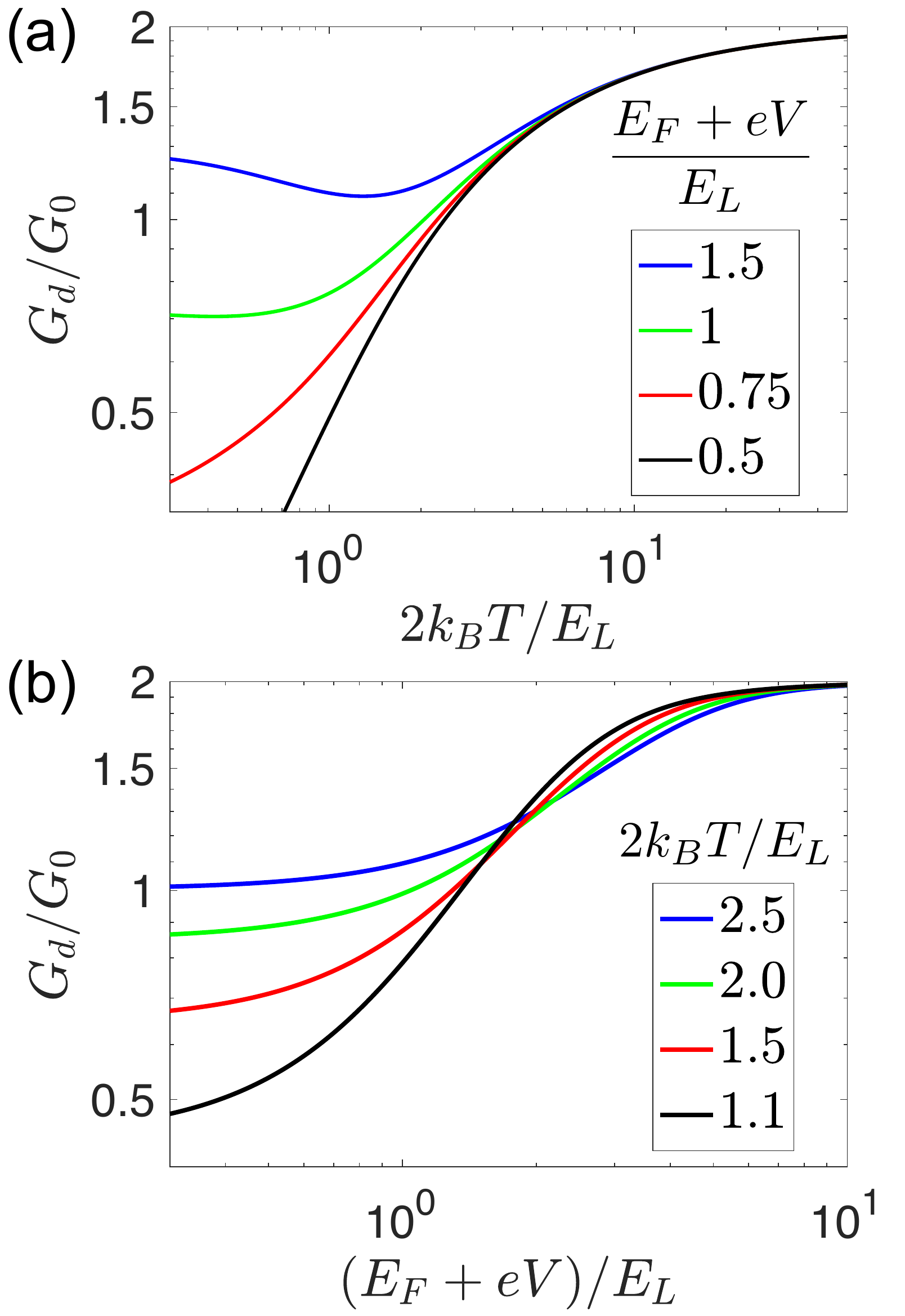}
\caption{(a) Differential conductance $G_d$ as a function of $T$ for $(E_F+eV)/E_L=:1.5, 1, 0.75, 0.5$.  (b) $G_d$ as a function of $V$ for $2k_BT/E_L=:2.5, 2, 1.5, 1.1$.}
\label{fig:diffcond}
\end{figure}

In order to study the full temperature dependence we introduce an energy scale $E_L$, which is defined in such a way that 
\begin{equation}
\ell(E_L) \equiv L, \ \textrm{ i.e.} \  E_L =  \sqrt{\frac{L \xi V_{\rm dis}^2 \Delta_{\rm edge}^2}{4 a \hbar^2 v^2}}.
\end{equation}
The differential conductance $G_d$, which depends on two parameters $(E_F+eV)/E_L$ and $2k_BT/E_L$,  is shown in Fig.~\ref{fig:diffcond}. In this analysis we have neglected the effects of electron-electron interactions beyond the mean field theory, and the energy and temperature dependence of the excitonic mean fields. Nevertheless, our results for the $G_d$ crossovers from a non-quantized to the  quantized value $G_d=2G_0$  with increasing voltage and temperature are in reasonable agreement with the experimental observations  \cite{Du-transport}. We consider the observations of these crossovers as very strong evidence of single-mode edge transport. 

\begin{figure}
\hspace{-0.2cm}\includegraphics[width=0.8 \linewidth]{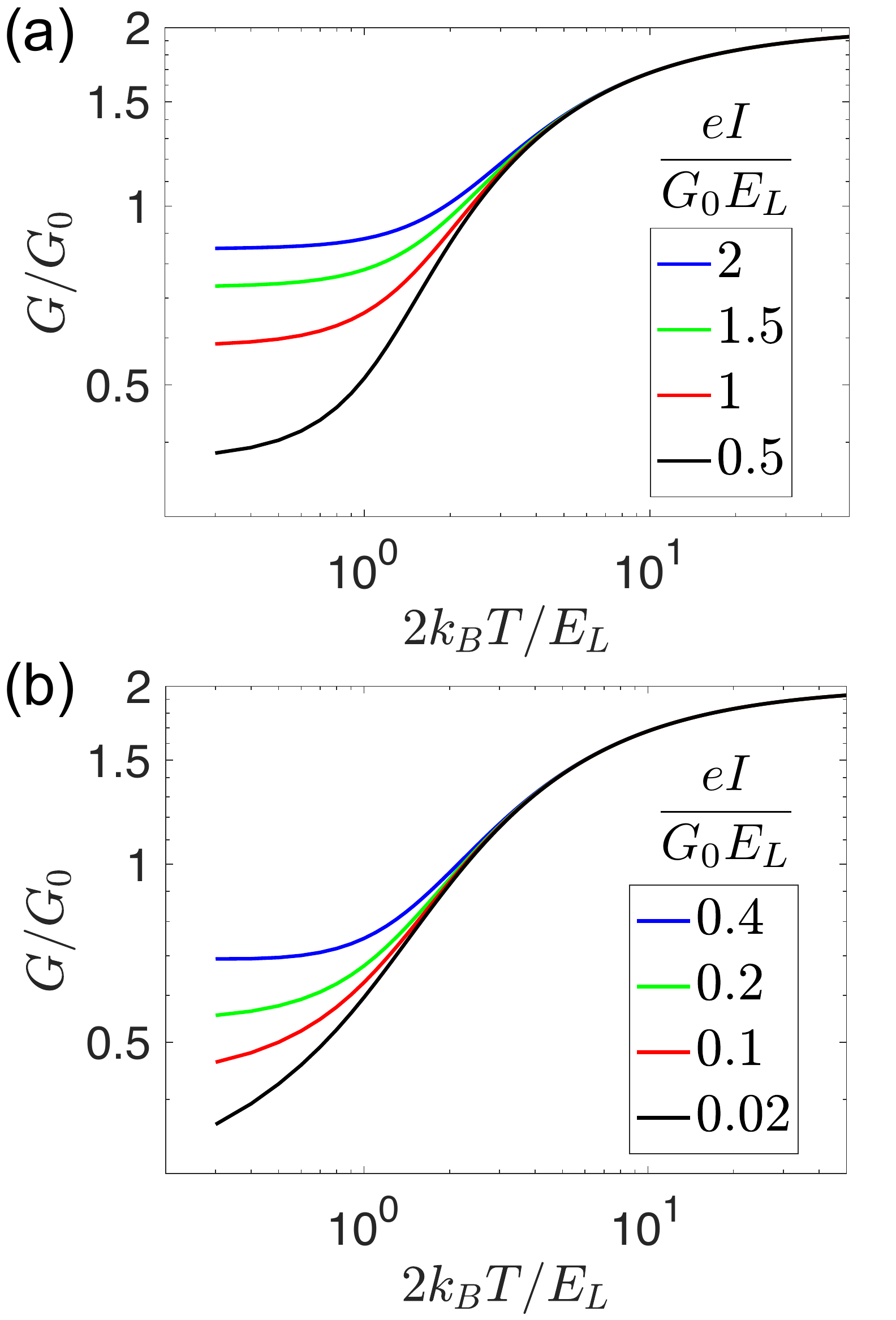}
\caption{(a) Conductance $G$ as a function of $T$ for $E_F=0$ and $eI/(G_0E_L)=:2, 1.5, 1, 0.5$. (b) Same for $E_F/E_L=0.7$ and $eI/(G_0E_L)=:0.4, 0.2, 0.1, 0.02$.}
\label{fig:cond_I}
\end{figure}

In the experiment \cite{Du-transport} the temperature dependence of the conductance 
\begin{equation}
G(E_F, V, T) =\frac{1}{V} \int_0^V dV G_d(E_F+eV,T) \label{G_eq}
\end{equation}
was reported also in a current $I$ biased situation. The theoretical predictions for this situation, obtained using Eqs.~(\ref{Gd_eq}), (\ref{ell_eq}), (\ref{G_eq}) and $I=GV$,  are shown in Fig.~\ref{fig:cond_I}. In this case, the shapes of the curves in the crossover regime depend on the Fermi energy $E_F$ and they resemble the experimental observations \cite{Du-transport} more in the case of  reasonably large values of $E_F$. In a more detailed microscopic description the crossing of the edge states may be buried within the bulk bands \cite{Pikulin18} so that reasonably large $E_F$ compared the energy of the crossing could naturally be realized in the experiments.

\section{Decoupling of bulk and edge transport in Corbino geometry}

We have shown that the transport experiments performed so far with InAs/GaSb devices are consistent with the system being in the TRS broken phase. However, it is difficult to rule out other possible theoretical explanations based on these experimental observations. In this Section, we propose a transport experiment, which could be used to proof the existence of the exotic TRS broken phase based on robust topological arguments. 

\begin{figure}
\centering
\vspace{0.1cm}
\includegraphics[width=0.88\columnwidth]{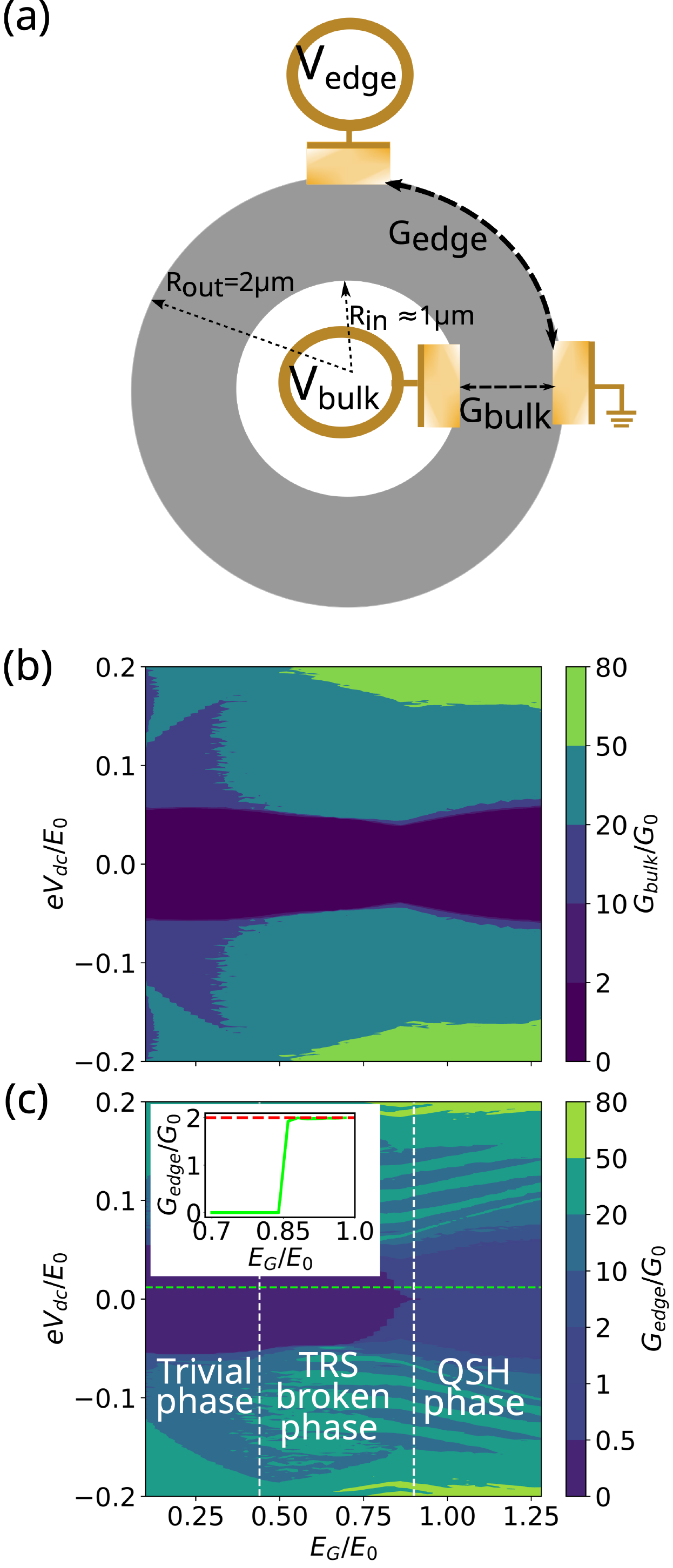}
 \caption{(a) Schematic illustration of a Corbino device and the transport paths corresponding to the bulk and edge differential conductances $G_{{\rm bulk}}$ and $G_{{\rm edge}}$.  The dimensions of the Corbino disc $R_{\rm in} \approx 1 \ \mu$m and $R_{\rm out} = 2 \ \mu$m are chosen so that the transport is (approximately) ballistic and the decay lengths of the evanescent bulk modes in the middle of the bulk gap are much shorter than the transport paths. (b),(c) $G_{{\rm bulk}}$ and $G_{{\rm edge}}$ as a function of $E_G$ and applied voltage $V_{dc}$. The inset in (c) shows $G_{{\rm edge}}$ as a function of $E_G$ (green line) for $eV_{dc}=0.012E_0$. The red dashed line is a guide to the eye.  The conductances have been calculated with the help of the tight-binding Hamiltonian which is generated from the continuum Hamiltonian,  defined by Eqs.~(\ref{eq:BHZ})-(\ref{p-wave-mean}), using the Kwant software package \cite{Groth_2014}.
 }
\label{fig:corbino-contour}
\end{figure}

Namely, we consider a Corbino device for decoupling the differential conductances corresponding to the bulk $G_{{\rm bulk}}$ and edge $G_{{\rm edge}}$ transport as illustrated in Fig.~\ref{fig:corbino-contour}. The dimensions of the Corbino disc $R_{\rm in} \approx 1 \ \mu$m and $R_{\rm out} = 2 \ \mu$m are chosen so that the transport is (approximately) ballistic and the decay lengths of the evanescent bulk modes in the middle of the bulk gap are much shorter than the transport paths. This guarantees that $G_{{\rm bulk}} \approx 0$ for the applied voltage satisfying $|e V_{dc}|<\Delta_{\rm bulk}/2$. Importantly, this allows to demonstrate that the transport gap does not close when the system is tuned from trivial to the QSH insulator phase by varying $E_G$ [see 
Fig.~\ref{fig:corbino-contour}(b)]. On the other hand, the edge conductance changes  smoothly from $G_{\rm edge} = 0$ (trivial phase) to $G_{\rm edge}=2G_0$ (QSH phase)  upon increasing $E_G$ demonstrating the closing of the edge  gap $\Delta_{\rm edge}$ at the transition to the QSH insulator phase [see 
Fig.~\ref{fig:corbino-contour}(c)]. Importantly, the bulk and edge conductances can be elegantly measured in the same device when the system is tuned {\it in-situ} from trivial to the QSH insulator phase using the gate voltages.   Such kind of experimental demonstration of a topological transition without a bulk transport gap closing would constitute a proof of the existence of TRS broken insulating phase.

\section{Conclusions and discussion}

We have discussed the possibility of unconventional topological phase transition between trivial and QSH insulator phases in band-inverted electron-hole bilayers. The hallmark of this transition is the existence of an intermediate insulating phase with spontaneously broken TRS. We have demonstrated that the transport properties of the system in the TRS broken phase are consistent with the observed transport characteristics of InAs/GaSb  devices, and we have shown that the measurement of the bulk and edge conductances in a Corbino device can provide unambiguous transport signatures of a topological transition without a bulk transport gap closing, proving the existence of the TRS broken phase.

Although we have focused on InAs/GaSb bilayers, we point out that band-inverted electron-hole systems can be realized in many semiconducting bilayers by creating a strong electric field at the barrier between the layers \cite{Sivan92,Kane94,Shapira99,Pohlt02,Seamons07}. In principle all these systems are potential candidates for supporting the interplay of excitonic correlations and the QSH effect, but for most of the semiconductors the barrier thickness may have to be so large that the hybridization gap between the electron and hole bands becomes too small to realize a sufficiently large topological gap in the QSH insulator phase. Our theory may also be applicable to HgTe bilayers \cite{Budich14}.

In a separate work \cite{Paul2}, we show also that in the presence of induced superconductivity the spontaneous time-reversal symmetry breaking allows to realize Majorana zero modes in the absence of magnetic field.

\begin{acknowledgments}
We thank D. I. Pikulin for useful discussions and comments.
The work is supported by the Foundation for Polish Science through the IRA Programme
co-financed by EU within SG OP.  We acknowledge
the computational resources provided by
the Aalto Science-IT project and the access to the computing facilities of the Interdisciplinary Center of Modeling at the University of Warsaw, Grant No. G87-1164 and G78-13.
\end{acknowledgments}

\appendix

\ \\ \

\section{Minimal model and mean field equations for excitonic correlations \label{APP:A}}

Based on the numerical solution of the Hartree-Fock mean field theory \cite{Pikulin14}, we know that the the main effect of intraband interactions (in the relevant part of the parameter space) is to renormalize the band structure. Therefore, we consider only the interband interactions
\begin{equation}
\hat{H}_I= - \sum_{s, s'} \sum_{\mathbf{k}, \mathbf{k}'} V_{\mathbf{k},\mathbf{k}'} c_{\mathbf{k}s1}^\dag  c_{\mathbf{k}s'2}  c_{\mathbf{k}'s'2}^\dag c_{\mathbf{k}'s1},
\end{equation}
where  $V_{\mathbf{k},\mathbf{k}'}$ describes the Coulomb interactions between the layers. 
On a mean-field level, the Hamiltonian is
\begin{equation}
\hat{H}_{\rm mf}=\hat{H}_0-\sum_{\mathbf{k}, s, s'} [ \Delta_{s s'} (\mathbf{k}) c_{\mathbf{k}s1}^\dag  c_{\mathbf{k}s'2}+h.c.],
\end{equation} 
where
\begin{equation}
\Delta_{s,s'}(\mathbf{k})=\sum_{ \mathbf{k}'}  V_{\mathbf{k},\mathbf{k}'} f_{s, s'}(\mathbf{k'}),
\end{equation}
\begin{equation}
f_{s, s'}(\mathbf{k}) \equiv \langle c_{\mathbf{k}s'2}^\dag c_{\mathbf{k}s1} \rangle=\sum_m n_F(E_{m\mathbf{k}}) \bigg[U_\mathbf{k} Q_{s s'} U^\dag_\mathbf{k} \bigg]_{mm},
\end{equation}
\begin{eqnarray}
Q_{\uparrow \uparrow}&=& \begin{pmatrix}
0  & 0 & 0 & 0\\
0 & 0 & 0 & 0 \\
1 & 0 & 0 & 0 \\
0 &  0 & 0 & 0
\end{pmatrix}, \ \ Q_{\uparrow \downarrow}= \begin{pmatrix}
0  & 0 & 0 & 0\\
0 & 0 & 0 & 0 \\
0 & 0 & 0 & 0 \\
1 &  0 & 0 & 0
\end{pmatrix}, \nonumber \\  Q_{\downarrow \uparrow}&=& \begin{pmatrix}
0  & 0 & 0 & 0\\
0 & 0 & 0 & 0 \\
0 & 1 & 0 & 0 \\
0 &  0 & 0 & 0
\end{pmatrix} , \ \ Q_{\downarrow \downarrow}= \begin{pmatrix}
0  & 0 & 0 & 0\\
0 & 0 & 0 & 0 \\
0 & 0 & 0 & 0 \\
0 &  1 & 0 & 0
\end{pmatrix}, 
\end{eqnarray}
$n_F(E)=\big(e^{E/(k_B T)}+1\big)^{-1}$ is the Fermi function, $T$ is the temperature and the transformation $U_\mathbf{k}$ diagonalizes 
\begin{equation}
{\rm diag}(E_{1\mathbf{k}}, E_{2 \mathbf{k}}, E_{3\mathbf{k}}, E_{4\mathbf{k}})=U_\mathbf{k} H_{\rm mf} (\mathbf{k}) U^\dag_\mathbf{k}
\end{equation}
the  mean field Hamiltonian 
\begin{widetext}
\begin{equation}
H_{\rm mf}(\mathbf{k})=\begin{pmatrix}
\xi(\mathbf{k})  & 0 & (A+\Delta_2)(k_x+ik_y) & -(\Delta_1+\Delta_z)\\
0 & \xi(\mathbf{k}) &(\Delta_1+\Delta_z) & -(A+\Delta_2)(k_x-ik_y) \\
(A+\Delta_2)^*(k_x-ik_y) & (\Delta_1+\Delta_z)^* & -\xi(\mathbf{k}) & 0 \\
-(\Delta_1+\Delta_z)^* &  -(A+\Delta_2)^*(k_x+ik_y) & 0 & -\xi(\mathbf{k}) 
\end{pmatrix}.
 \end{equation}
 \end{widetext}
Here we have utilized the fact that 
the excitonic mean field can be approximated as
\begin{equation}
\Delta^{\rm mf}=i \Delta_1 \sigma_2-\Delta_2 (k_x \sigma_3+ik_y\sigma_0),  \label{mfsolution}
\end{equation}
where $\Delta_1$ and $\Delta_2$ are complex bosonic fields describing $s$-wave and $p$-wave excitonic correlations, respectively.

By inverting the interaction matrix and substituting the ansatz (\ref{mfsolution}) to the mean field equation, we obtain
\begin{equation}
\frac{d_0^2}{L^2}\sum_{\mathbf{k}}[ f_{\uparrow, \downarrow}(\mathbf{k})-f_{\downarrow, \uparrow}(\mathbf{k})]=2 \frac{d_0^2}{L^2} \sum_{\mathbf{k}, \mathbf{k}'} V^{-1}_{\mathbf{k} \mathbf{k}'} \Delta_1 =\frac{1}{g_s} \Delta_1
\end{equation}
and
\begin{eqnarray}
&& \frac{d_0^2}{L^2}\sum_{\mathbf{k}}[ -f_{\uparrow, \uparrow}(\mathbf{k})(k_x-ik_y)+f_{\downarrow, \downarrow}(\mathbf{k}) (k_x+ik_y)] \nonumber \\ &=& 2 \frac{d_0^2}{L^2} \sum_{k, k'} V^{-1}_{\mathbf{k},\mathbf{k}'} \Delta_2 (k_x k_x'+k_y k_y')=  \frac{1}{g_p d_0^2}  \Delta_2,
\end{eqnarray}
where we have defined effective interaction strengths $g_s$ and $g_p$ for the $s$-wave and $p$-wave excitonic correlations as 
\begin{equation}
g_s^{-1}=2 \frac{d_0^2}{L^2} \sum_{k, k'} V^{-1}_{\mathbf{k},\mathbf{k}'}, \ \ \
g_p^{-1}= 2 \frac{d_0^4}{L^2} \sum_{k, k'} V^{-1}_{\mathbf{k},\mathbf{k}'} (k_x k_x'+k_y k_y').
\end{equation}
The length scale $d_0$ is introduced to guarantee that the interaction strengths have a unit of energy, and it can in principle be chosen arbitrarily. However, we know that in the case of Coulomb interaction the natural length $d_0$ and energy $E_0$ scales are determined so that the kinetic and interaction energies are equal
\begin{equation}
E_0=\frac{\hbar^2}{ d_0^2} \frac{1}{m}=\frac{1}{4\pi \epsilon \epsilon_0} \frac{e^2}{d_0}.
\end{equation}
This way we obtain the mean field equations (\ref{s-wave-mean}) and (\ref{p-wave-mean}) given in the main text.

\bibliography{bibliography}

\end{document}